\documentclass[fleqn,usenatbib]{mnras}
\usepackage{newtxtext,newtxmath}
\usepackage[T1]{fontenc}
\usepackage{units} 
\usepackage{lineno}
\usepackage{color}
\usepackage{epsfig} 
\usepackage{float}
\usepackage{bm}
\DeclareRobustCommand{\VAN}[3]{#2} 
\let\VANthebibliography\thebibliography
\def\thebibliography{\DeclareRobustCommand{\VAN}[3]{##3}\VANthebibliography}
\usepackage{graphicx}
\usepackage{amsmath}     

\usepackage{amssymb}
\usepackage{amsfonts}
\usepackage{caption}
\usepackage{subfigure} 
\usepackage{multirow}
\usepackage[figuresright]{rotating}
\usepackage{bm}
\usepackage{longtable} 
\usepackage{booktabs}
\usepackage{multirow} 
\usepackage[mathscr]{euscript} 
\usepackage{soul}
\usepackage{listings}

\newcommand\be{\begin{eqnarray}}
\newcommand\ee{\end{eqnarray}}

\newcommand{\fraction}[2]{\left(\frac{#1}{#2}\right)}

\title[Scatter in the Relation between PRS Luminosity and FRB RM]{Scatter in the Relation between Persistent Radio Source Luminosity and Fast Radio Burst Rotation Measure: A Window into Circum-burst Environments}

\author[Yang]{
Yuan-Pei Yang$^{1}$\thanks{E-mail: ypyang@ynu.edu.cn (YPY)}\\
$^{1}$South-Western Institute for Astronomy Research, Yunnan Key Laboratory of Survey Science, Yunnan University, Kunming, Yunnan 650504, People's Republic of China\\
}  

\date{Accepted XXX. Received YYY; in original form ZZZ}

\pubyear{2026}

\begin{document}

\label{firstpage}
\pagerange{\pageref{firstpage}--\pageref{lastpage}}
\maketitle

\begin{abstract}

The association of persistent radio sources (PRSs) with repeating fast radio bursts (FRBs) offers unique insights into their circum-burst environments. Building upon the physical link between PRS luminosity ($L_\nu$) and FRB rotation measure (RM), we introduce a novel diagnostic framework utilizing the intrinsic scatter of the $L_\nu - |{\rm RM}|$ relation as a physical probe of nebula dynamics. We show that this scatter encodes critical information regarding the temporal evolution of the nebula radius ($R \propto t^\alpha$). By deriving a generic scaling $L_\nu \propto R^\epsilon |{\rm RM}|$, we apply this framework to the current sample of five confirmed FRB-PRS systems as a proof-of-concept. Our preliminary analysis yields a combined evolutionary index of $\alpha|\epsilon| = 1.5 \pm 0.8$. This measurement illustrates the potential of our methodology to distinguish among diverse astrophysical scenarios, such as supernova remnants (SNRs) in different evolutionary stages, pulsar wind nebulae (PWNe) driven by constant wind or varying winds, and binary bow-shock systems. While the current conclusions are limited by the small sample size, this work establishes a robust statistical methodology for future population studies. As next-generation radio surveys localize a larger population of active repeaters, this framework will serve as a powerful tool to statistically constrain the physical origin and dynamical life cycle of FRB circum-burst environments.

\end{abstract}


\section{Introduction}

Fast radio bursts (FRBs) are millisecond-duration radio transients at cosmological distances, characterized by extreme brightness temperatures that necessitate coherent radiation mechanisms. Up to the present, more than 4500 FRB sources have been detected and nearly 100 have been confirmed to be repeaters \citep{CHIME21,CHIME26}. 
The association between FRB 20200428 and the Galactic magnetar SGR J1935+2154 provides compelling evidence that at least a subset of FRBs can originate from magnetars \citep{CHIME20,Bochenek20,Mereghetti20,Li21,Ridnaia21,Tavani21}. In contrast, the precise localization of FRB 20200120E to an exceptionally old globular cluster \citep{Bhardwaj21,Kirsten22} challenges scenarios involving young magnetars formed through core-collapse supernovae. Instead, this finding favors progenitor channels linked to older stellar populations, such as compact binary systems or alternative evolutionary pathways \citep{Wang16,Zhang20,Kremer21,Lu22}. 
These contrasting cases highlight the diversity of FRB progenitors and underscore the importance of identifying multiwavelength counterparts for extragalactic FRBs  (see the recent reviews of \citet{Zhang23,Zhang24}). To date, apart from host galaxies, persistent radio sources (PRSs) remain the only confirmed multiwavelength counterparts of extragalactic FRBs. These PRSs as compact sources near the FRB sources are physically distinct from the star-formation activity of their host galaxies and provide a unique window into the immediate circum-burst medium and the central engine of these FRB repeaters.

At present, five repeating FRBs have been securely associated with such compact PRSs \citep{Chatterjee17,Niu22,Bruni24,Bruni25,Bhusare25,Zhang25,Ibik24,Moroianu26}. High-resolution radio observations have revealed a remarkable diversity in their properties, spanning a wide range of specific luminosities ($L_{\nu}$), spectral characteristics, and the Faraday rotation measures (RMs) of their associated bursts. Their key properties are summarized as follows: 
\begin{itemize}
    \item \textbf{FRB 20121102A:} As the prototype of this class residing in a dwarf galaxy at $z=0.19273$ \citep{Tendulkar17}, its associated PRS is physically compact (with a projected size $<0.7$ pc, \citet{Marcote17}) and highly luminous, with an observed peak specific luminosity of $L_{\nu,{\rm peak}} \simeq 2.9\times 10^{29}~{\rm erg~s^{-1}Hz^{-1}}$ at 1.6 GHz band \citep{Chatterjee17}. The bursts exhibit an exceptionally high and decreasing RM of $|{\rm RM}| \sim 10^4-10^5~{\rm rad~m^{-2}}$ \citep{Michilli18,Hilmarsson21} and a complex evolution of DM with a structural change \citep{Wang25}, indicating an evolving extreme magneto-ionic environment. Significant flux variability was detected in this PRS covering the longest timescale recorded to date, with consistent amplitudes across short and long timescales \citep{Yang24}. 
    
    \item \textbf{FRB 20190520B:} Often considered a ``twin'' to FRB 20121102A, this source is co-located with a compact PRS (with a projected size $<9$ pc as reported by \citet{Bhandari23}) in a dwarf host galaxy at $z=0.241$ \citep{Niu22}. The PRS luminosity is similarly high (with an observed peak specific luminosity of $L_{\nu,{\rm peak}} \simeq 4.9 \times 10^{29}~{\rm erg~s^{-1}Hz^{-1}}$ at 1.5 GHz band). Notably, it exhibits a significantly decreasing host DM contribution ($\rm DM_{\rm host} \simeq 900~{\rm pc~cm^{-3}}$) with a decay rate of $d{\rm DM}/dt=-12.4~{\rm pc~cm^{-3}yr^{-1}}$ \citep{Niu26}. Meanwhile, its highly variable RM showed dramatic sign reversals from $-36,000$ to $+20,000~{\rm rad~m^{-2}}$ \citep{Anna-Thomas23}, suggesting a turbulent, dense environment \citep{Yang23} or a binary system \citep{Wang22,Zhang25c}. The variability feature of the PRS light curve is similar to that associated with FRB 20121102A \citep{Yang24}. 
    
    \item \textbf{FRB 20201124A:} Unlike the previous two, this source resides in a massive spiral galaxy at $z=0.0979$ \citep{Fong21}. Deep VLA observations resolved a compact PRS (with a projected size $<0.7$ kpc) distinct from the host's diffuse star formation. It is significantly fainter (with an observed peak specific luminosity of $L_{\nu,{\rm peak}} \simeq 7.9\times10^{27}~{\rm erg~s^{-1}Hz^{-1}}$ at 22 GHz) and exhibits an inverted spectrum of $F_\nu \propto \nu^{1.0}$ \citep{Bruni24}, in contrast to the previous two. Its RM is comparatively lower, with values varying between $-365$ and $-890$ rad m$^{-2}$ during the first 36 days of observations, followed by a stable, constant RM phase \citep{Xu22}.
    
    \item \textbf{FRB 20240114A:} This active repeater exhibited unprecedented hyper-activity with over 11,000 bursts in 214 days, challenging standard magnetar models \citep{Zhang25b}, and it is also associated with a PRS (with a projected size $<4$ pc) in a sub-solar metallicity dwarf starburst galaxy at $z=0.13056$ \citep{Bruni25,Bhusare25,Zhang25}. The PRS has an observed peak luminosity of $L_{\nu,{\rm peak}} \simeq 3.5\times10^{28}~{\rm erg~s^{-1}Hz^{-1}}$ at 1.3 GHz \citep{Zhang25} and an RM of $338~{\rm rad~m^{-2}}$ \citep{Tian24}.  
    
    \item \textbf{FRB 20190417A:} Recently localized with milliarcsecond precision (with a projected size $<23$ pc) at a star-forming galaxy at $z = 0.12817$, this source represents an extreme case with an exceeding DM contribution \citep{Moroianu26}, like that of FRB 20190520B \citep{Niu22}. It is associated with a compact PRS (with an observed specific luminosity $L_{\nu,{\rm peak}} \simeq 8.7 \times 10^{28}~{\rm erg~s^{-1}Hz^{-1}}$ at 1.4 GHz and possesses the largest known lower limit on the host DM contribution ($>1228$ pc cm$^{-3}$). Its RM is high and variable ($+4000$ to $+5000~{\rm rad~m^{-2}}$), placing it in an intermediate regime between the prototypes and the fainter sources. 
\end{itemize}
Despite their diverse host environments and spectral behaviors, these five systems collectively exhibit a broad, yet constrained, distribution in their specific luminosities ($L_\nu$) and Faraday rotation measures (RMs).
This growing sample suggests that PRSs may be a common feature of the repeating FRB population, rather than rare or exceptional associations. 
However, no PRS has been linked to a one-off FRB to date, implying a possible diversity in FRB progenitors or distinct emission mechanisms \citep{An25}.

The nature of this nebula contributing PRS and RM remains a subject of open debate. 
Various possible scenarios have been proposed in the literature.
\citet{Yang16} first proposed that an FRB source embedded in a self-absorbed synchrotron nebula can heat nebular electrons through absorption process. The resulting hardened electron spectrum produces a spectral hump near the synchrotron self-absorption frequency, providing an observable signature of an embedded FRB. This model accounts for the observed special spectrum of the PRS associated with FRB 20121102A \citep{Li20}.
The magnetar wind nebula scenario suggests that the PRS can originate from a young millisecond magnetar formed in a supernova or gamma-ray burst, powered by synchrotron emission from a magnetized wind nebula either through shocks interacting with the supernova ejecta \citep{Murase16,Metzger17,Margalit18,Rahaman25} or directly with the surrounding interstellar medium \citep{Dai17,Yang19}.
Alternatively, the ``hypernebula'' model proposed that a PRS can be produced by a nebula of relativistic electrons heated at the termination shock of super-Eddington disk winds and jets from a black hole or neutron star during a brief unstable mass-transfer phase in a binary system \citep{Sridhar22}.
Finally, accreting wandering massive black holes in dwarf galaxies represent another plausible origin for PRSs \citep{Eftekhari20,Reines20,Dong24}.

While the precise origin has yet to be determined, a physical connection between the burst source and the PRS can be generally inferred from the concurrent requirements for particle acceleration and Faraday rotation within a magnetized medium \citep{Yang20,Yang22}, also see Figure \ref{diagram}. First, the observed non-thermal continuum spectra of PRSs indicate the presence of relativistic electrons, which must be accelerated by energetic processes such as shocks or magnetized pair winds. Second, the large RMs imply that the radio bursts propagate through a dense, highly magnetized plasma. If the persistent synchrotron emission and the large Faraday rotation originate from the same physical volume, essentially a magnetized nebula surrounding the central engine, a generic correlation is expected between the specific luminosity of the PRS and the rotation measure ($L_{\nu} \propto |{\rm RM}|$) as proposed by \citet{Yang20,Yang22}. In this framework, a more luminous PRS, driven by higher energy injection or stronger magnetic fields, naturally corresponds to a denser magneto-ionic environment and, consequently, a larger $|{\rm RM}|$. Such a ``$L_\nu-|{\rm RM}|$ relation'' has been confirmed by an increasing number of detected PRSs \citep{Bruni24,Bruni25,Ibik24,Moroianu26} and been applied to study the FRB cosmology \citep{Zhang25d,Gao25}.

This generic nebular model predicts a theoretical relation between PRS luminosity and FRB RM. However, observational data show a significant intrinsic scatter in the $L_{\nu}-|{\rm RM}|$ plane. This scatter cannot be explained solely by measurement uncertainties, and it is likely physical, encoding vital information regarding the structural diversity of the circum-burst environments. In this work, we investigate the deviations of individual sources from the mean theoretical scaling. We demonstrate that this scatter provides a powerful diagnostic tool to constrain the physical radius distribution of the emitting nebulae and further infer the long-term evolution of the nebulae. 
The paper is organized as follows. In Section \ref{sec2}, we briefly summarize and further refine the theoretical framework of the $L_\nu-|{\rm RM}|$ relation. In Section \ref{sec3}, we analyze how the scatter of the $L_\nu-|{\rm RM}|$ relation reflect the distribution of the PRS-RM nebulae. In Section \ref{sec4}, we discuss the constraint on various candidate astrophysical scenarios that generate PRS-RM. Finally, the results are discussed and summarized in Section \ref{sec5}. Throughout this paper, the mathematical symbol ``$\log$'' refers to the logarithm to base 10, and ``$\ln$'' refers to the natural logarithm.
The $\Lambda$CDM cosmological parameters in this work are taken as $\Omega_{\rm m}=0.315$ and $H_0=67.36~{\rm km~s^{-1}~Mpc^{-1}}$ \citep{Planck20}.

\section{Theoretical Framework of PRS Luminosity -- FRB RM relation}\label{sec2}

\begin{figure}
    \centering
    \includegraphics[width = 1.0\linewidth, trim = 30 30 30 30, clip]{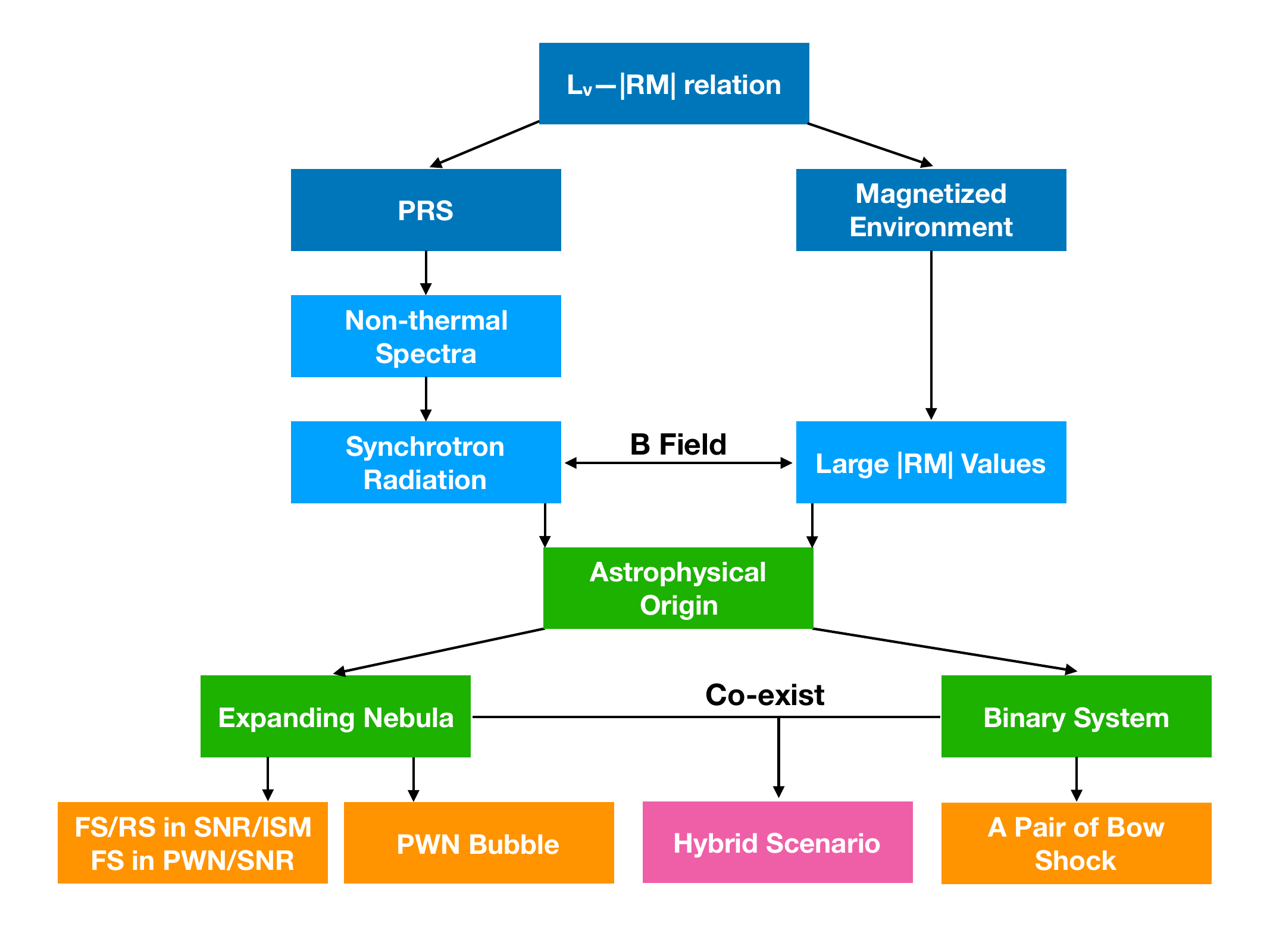} 
    \caption{A flowchart showing the physical connections between PRS and RM. ``FS'' and ``RS'' denote forward shock and reverse shock, respectively. 
    We emphasize that, in a PWN/SNR system, the radio emission originates from fossil electrons within the PWN bubble rather than the termination shock. 
    }\label{diagram}
\end{figure}

\begin{figure*}
    \centering
    \includegraphics[width = 0.49\linewidth, trim = 0 0 0 0, clip]{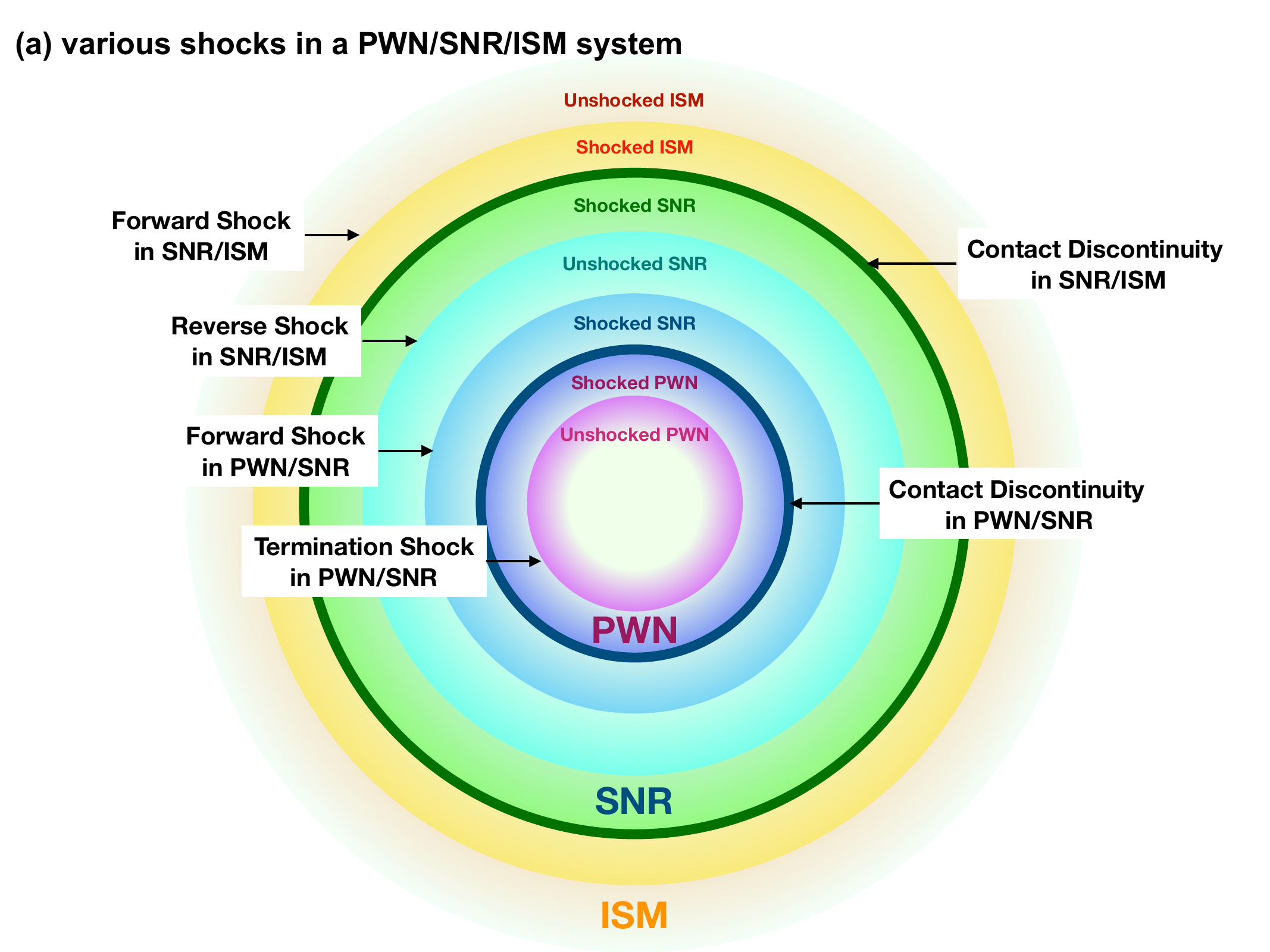} 
    \includegraphics[width = 0.49\linewidth, trim = 0 0 0 0, clip]{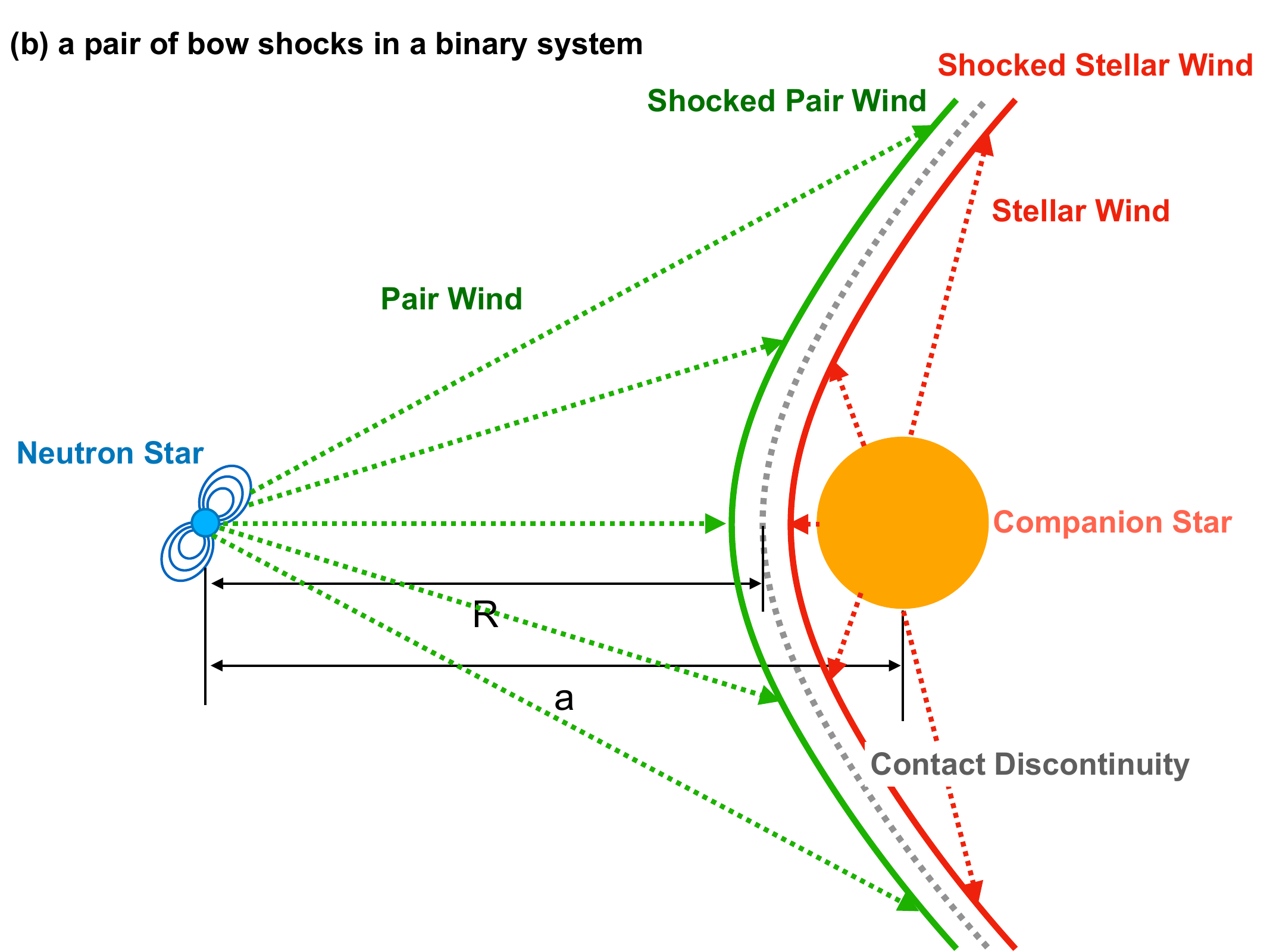} 
    \caption{Various shocks in a PWN/SNR/ISM nebula (panel a) and a pair of bow shocks in a binary system (panel b).
    }\label{shocks}
\end{figure*}

\citet{Yang20,Yang22} suggested that the large RM of an FRB repeater and its associated PRS may originate from the same physical region. Under this assumption, a generic relation can be established between the $|{\rm RM}|$ value and the PRS specific luminosity, which depends only weakly on the specific PRS model.
Considering that the relativistic electrons that emit non-thermal PRS should be from shocks or magnetized particle winds, the possible astrophysical scenarios can involve forward shock (FS) / reverse shock (RS) in a supernova remnant (SNR) / interstellar medium (ISM) system, FS / termination shock (TS) in a pulsar wind nebula (PWN) / SNR system, pulsar-wind/companion-wind bow shocks in a binary system. 
A flowchart is summarized in Figure \ref{diagram} and the corresponding schematic diagram is shown in Figure \ref{shocks}.
In all these scenarios, the bulk motion of the downstream and the contact discontinuity is usually non-relativistic in the observer frame\footnote{Relativistic bulk motion of the downstream primarily occurs in astrophysical contexts such as gamma-ray bursts and active galactic nuclei; however, in both cases, it is highly unlikely to be related to the origin of FRBs.}, while the shock velocity relative to the unshocked medium can be either relativistic (e.g., TS in PWN, pair-wind bow shock in a binary system) or non-relativistic (e.g., the remaining scenarios mentioned above). 
The PRS-RM region can contain non-relativistic and/or relativistic electrons. We define the momentum distribution of all these electrons as $n_e(\hat p)=n_{e,0}f(\hat p)$, where $\hat p\equiv p/m_ec=\sqrt{\gamma^2-1}$ is the dimensionless momentum and $\hat p\simeq\gamma$ for $\gamma\gg1$, $p$ is the dimensional momentum, $\gamma$ is the Lorentz factor of electrons, 
$f(\hat p)=n_e(\hat p)/n_{e,0}$ with $\int f(\hat p)d\hat p=1$ and $n_{e,0}$ is the total electron number density.
In a relativistic context, the inertia of the electron increases, leading to an effective mass of $m_e^*\sim\gamma m_e$. Thus, in the PRS-RM region, the RM can be approximately written as \footnote{In one specific treatment, accounting for the anisotropy of electron momentum in a magnetic field and assuming an ultra-relativistic Maxwellian distribution, a logarithmic term in the Lorentz factor, $\ln \gamma/2$, appears in the integrand \citep{Quataert00}. Since $\ln \gamma/2$ is of order unity and its specific form derives from the ultra-relativistic Maxwellian distribution (which does not include the power-law component at high-energy band), this logarithmic term is neglected in the present work. This approximation has a negligible impact on the final results.}
\begin{align} 
{\rm RM}\simeq\frac{e^3}{2\pi m_e^2c^4}\int_0^D ds~\frac{n_{e,0}B_{\parallel}}{\gamma_c^2},\label{RM}
\end{align}
and the critical electron Lorentz factor is defined by
\begin{align} 
\gamma_c\equiv\left(\int_0^{\infty}d\hat p\frac{f(\hat p)}{1+\hat p^2}\right)^{-1/2}.\label{gammac}
\end{align}  
where $\gamma=\sqrt{1+\hat p^2}$ is used in the above formula.
For the non-relativistic regime, the RM is given by the classical formula due to $\gamma_c\sim1$; while for the relativistic regime, the RM is suppressed by a factor of $\sim\gamma_c^2$. As shown in Eq.(\ref{RM}), the properties of the electron momentum distribution play a key role in determining the magnitude of the RM. In particular, the large $|{\rm RM}|$s observed in some repeating FRBs suggest that low-energy electrons dominate in the region responsible for the RM.

In the distribution of $f(\hat p)$, electrons above a certain momentum of $\hat p_{\rm obs}$ will produce the observed GHz wave via synchrotron radiation.
We consider the synchrotron radiation from relativistic electrons under the magnetic field $B$ in the PRS-RM region.
The radiation power for a single electron is $P=(4/3)\sigma_{\rm T}c\gamma^2 B^2/8\pi$, and the characteristic frequency is $\nu=\gamma^2eB/2\pi m_ec$.
Thus, the specific power is $P_\nu\simeq P/\nu=m_ec^2\sigma_{\rm T}B/3e$ and it is independent of $\nu$. For a given observed frequency $\nu\sim1~{\rm GHz}$, the corresponding electrons are required to have a momentum of 
\begin{align}
\hat p_{\rm obs}\simeq \gamma_{\rm obs}\sim\left(\frac{2\pi m_ec\nu}{eB}\right)^{1/2}\simeq600\left(\frac{\nu}{1~{\rm GHz}}\right)^{1/2}\left(\frac{B}{1~{\rm mG}}\right)^{-1/2}.\label{gammaghz}
\end{align} 
Since the bulk motion of the PRS-RM region should be non-relativistic for the possible scenarios we are interested here, e.g., PWN, SNR and binary system, the emission frequency at the bulk motion frame is approximately equal to that in the observer frame.
We further define $\zeta_e$ as the fraction of electrons that radiate synchrotron emission in the GHz band, then the fraction $\zeta_e$ is given by $\zeta_e\equiv F(>\hat p_{\rm obs})$,
\begin{align} 
\zeta_e\equiv F(>\hat p_{\rm obs})=\int_{\hat p_{\rm obs}}^{\infty}d\hat pf(\hat p),
\end{align} 
where $F(>\hat p)$ is the cumulative distribution function (CDF) associated with the probability density function (PDF) $f(\hat p)$.
We consider that the PRS-RM region has a radius of $R$ and a thickness of $\Delta R$, then total number of relativistic electrons that emits GHz wave is approximately $N_e\simeq4\pi R^2\Delta R\zeta_e n_e/3$ for $\Delta R\sim R$. 
The specific luminosity of synchrotron radiation is given by $L_{\nu}=N_e P_\nu$ \citep{Yang20,Yang22}
\begin{align} 
L_{\nu}&=\frac{64\pi^3}{27}\zeta_e\gamma_{c}^2m_ec^2R^2\left|{\rm RM}\right|\simeq5.7\times10^{29}~{\rm erg~s^{-1}~Hz^{-1}} \nonumber \\
&\times \left(\frac{\zeta_e\gamma_{c}^2}{0.01}\right) \left(\frac{R}{1~{\rm pc}}\right)^2 \left(\frac{\left|{\rm RM}\right|}{10^3~{\rm rad~m^{-2}}}\right). \label{lum}
\end{align} 
where the factor of $\zeta_e\gamma_c^2$ only depends the electron momentum distribution and the observed frequency, and it can be calculated by
\begin{align}  
\zeta_e\gamma_c^2=\left(\int_0^{\infty}d\hat p\frac{f(\hat p)}{1+\hat p^2}\right)^{-1}\int_{\hat p_{\rm obs}}^{\infty}d\hat pf(\hat p).\label{zg2}
\end{align}  
The above PRS luminosity--FRB RM relation $L_\nu\propto \zeta_e\gamma_c^2R^2|{\rm RM}|$ in Eq.(\ref{lum}) has the following implications and considerations:
\begin{itemize}
\item In the $L_\nu-|{\rm RM}|$ relation, since the RM is contributed only by the line-of-sight component of the magnetic field, whereas the PRS depends on the total magnetic field strength, we emphasize that, in practice, it is suggested to be taken to be the maximum observed value of $|{\rm RM}|$ as an estimator.

\item Due to the large RM uncertainty of the host interstellar medium, the data with $|{\rm RM}|\lesssim10^2-10^3~{\rm rad~m^{-2}}$ might substantially deviate from the predicted relation. 
However, for a sufficiently large sample in the future, one can mainly focus on the sample with $|{\rm RM}|\gtrsim10^2-10^3~{\rm rad~m^{-2}}$ (as appeared in the current PRS sample). In this case, the host RM contribution can be neglected.

\item $|{\rm RM}|$ can be a function of the nebula radius $R$ in the proposed nebula models \citep[e.g.,][]{Yang16,Murase16,Metzger17,Margalit18}, but it also depends on other physical quantities, such as the magnetic field, gas density, the initial mass and kinetic energy of the SN ejecta, etc. In this case, the correlation between $L_\nu$ and $|{\rm RM}|$ can even deviate from a power-law index of 1 more or less. However, as we will prove as below, one can still fit the observed data with $L_\nu\propto |{\rm RM}|$ and the corresponding scatter will reflect the distribution of the nebula radius $R$ in this case.
\end{itemize}

\section{Implication of Scatter in $L_\nu-|{\rm RM}|$ relation}\label{sec3}

We have demonstrated that, for a given observed frequency (e.g., $\nu \sim 1,\mathrm{GHz}$), the factor $\zeta_e \gamma_c^2$ is primarily determined by the electron distribution function $f(\hat p)$.
In an expanding nebula, e.g., SNR or PWN, the electron distribution $f(\hat p)$ changes as the nebula expands. As a result, $\zeta_e \gamma_c^2$ becomes largely dependent on the nebular radius $R$ (see Section~\ref{sec4} for details).
This implies that, for an expanding nebula, $\zeta_e \gamma_c^2$ can be treated as a function of the nebular radius $R$. Accordingly, the above $L_\nu$–$|{\rm RM}|$ relation can be written in a more general form as 
\begin{align}  
L_\nu\propto R^{\epsilon}|{\rm RM}|.\label{lum2} 
\end{align}  
where $\epsilon$ is a power-law index.
\emph{In this case, the distribution of the nebula radius leads to the scatter of the $L_\nu-{|\rm RM|}$ relation.}

\subsection{Methodology}\label{methodology}

\begin{table*}
\centering
\caption{Radius Evolution of various shocks and bubbles in a PWN/SNR/ISM system}\label{table} 
\begin{tabular}{llcccc}
\toprule
& & FS in SNR/ISM & RS in SNR/ISM & FS in PWN/SNR & PWN Bubble \\
\midrule
\multicolumn{2}{l}{Free-Expansion} & $R\propto t$, $V_{\rm sh}\sim$const. & $R\sim t$, $V_{\rm sh}\propto t^{3/2}$ & $R\propto t^{6/5}$, $V_{\rm sh}\propto t^{1/5}$ & $R\propto t^{6/5}$ \\
\midrule
\multirow{2}{*}{Sedov-Taylor} 
& Constant Wind & $R\propto t^{2/5}$, $V_{\rm sh}\propto t^{-3/5}$ & Shock absent &  Shock absent & $R\propto t^{11/15}$\\
& Decaying Wind & $R\propto t^{2/5}$, $V_{\rm sh}\propto t^{-3/5}$ & Shock absent &  Shock absent & $R\propto t^{3/10}$\\
\bottomrule
\end{tabular}
\\
\vspace{0.3em}
{\flushleft
\footnotesize{{\bf Notes:} 
``FS'' and ``RS'' denotes forward shock and reverse shock, respectively. $R$ is the shock/bubble radius and $V_{\rm sh}$ is the shock velocity related to the upstream.
During the Sedov–Taylor phase, the RS in the SNR/ISM system has completely swept through all of the SN ejecta, and the shock radius approaches the center of the SNR.
Since this RS reheats the SNR ejecta, the sound velocity increases by a large factor, which causes that the FS in the PWN/SNR system has also been absent \citep{Swaluw01}.
For the RS in the SNR/ISM system, the shock velocity related to the upstream is $V_{\rm sh}=R/t-dR/dt$. Due to the contribution of an additional second-order small quantity \citep{McKee95}, a time-dependent shock velocity is obtained.
In particular, for the PWN bubble scenario, the radio radiation is emitted by fossil electrons in the bubble. In this case, $R$ denotes the bubble radius (i.e., the contact discontinuity) and the shock velocity is omitted here. The derivation of the scaling laws listed in the table can be found in the Appendix \ref{appendix}.
\\
}}
\end{table*}

First, we derive how the observed scatter in the $L_\nu-|{\rm RM}|$ relation constrains the evolution of nebular radius.
In most scenarios of interest, the distribution of nebular radii is primarily determined by their ages\footnote{Note that in the bow shock scenario within a binary system, the bow-shock nebula radius does not evolve with time and is instead primarily determined by the orbital separation (see Section \ref{binary}). This contrasts with expanding nebulae, whose sizes increase progressively as they age.}. 
We first generally consider the time evolution of the nebula radius as
\begin{align} 
R\propto t^{\alpha}, 
\end{align} 
where $\alpha$ is the temporal index of the nebula radius evolution.
Such a power-law evolution is consistent with the scenarios of PWN and SNR \citep{Reynolds84,McKee95,Swaluw01}, also see Table \ref{table} in a summary. 
We assume that the formation rate of the PRS sources is steady, in which case, the source number is proportional to the source age. Thus, the CDF of the nebula radius is given by
\begin{align}  
F(<R)
\propto R^{1/\alpha}.\label{CMFR}
\end{align}  
Using the definition of $x\equiv\ln R$, the above CDF can be written as $F(<x)\propto e^{x/\alpha}$.
Assuming the range of $R$ is sufficiently large, $\ln(R_{\max}/R_{\min})\gg\alpha$, where $R_{\max}$ and $R_{\min}$ are the maximum and minimum nebula radius, respectively, the truncated exponential distribution can be approximated by a semi-infinite distribution. Under this assumption, the standard deviation of $x$ is given by $\sigma_x\simeq\alpha$, i.e., $\sigma_{\ln R}\simeq\alpha$. Converting this back to the base-10 logarithm using the change of base formula $\log R=\ln R/\ln 10$, one finally has 
\begin{align}  
\sigma_{\log R}\simeq\frac{\alpha}{\ln 10}.\label{sigmalogR}
\end{align}   
Therefore, once $\sigma_{\log R}$ is measured, we will know how the nebula radii evolve over time. 

\begin{table*}
\centering
\caption{FRB-PRS sample considered in this work}\label{table1}
\begin{tabular}{ccccccccc}
    \hline
FRB Name  &  $z^{\rm a}$    &   $d_{\rm{L}}^{\rm b}$  & ${\rm RM}^{\rm c}$ & $F_{\nu}^{\rm d}$\ & $\nu^{\rm e}$ & $L_{\nu}^{\rm f}$  & References$^{\rm g}$\\
&       &   ($\unit{Gpc}$)  &  ($\unit{rad\,m^{-2}}$)  & ($\mu\unit{Jy}$)   &  ($\unit{GHz}$) & ($10^{29}\unit{erg\,s^{-1}Hz^{-1}}$)  &  \\
    \hline
FRB  20121102A &   $0.19273$    &   $0.98$  &  $1.46\times10^{5}$  &   203  & $6$ & $1.9$  & 1,2,3,4\\

FRB 20190520B    & $0.241$  & $1.25$ & $-3.6\times10^{4}$ & $152$ & $5.5$ & $2.3$ & 5,6\\

FRB 20201124A    & $0.0979$  & $0.47$ & $-889.5$ & $8.2$ & $6$ & $0.020$ & 7,8,9\\

FRB 20240114A    & $0.13056$  & $0.64$ & $338.1$ & $46$ & $5$ & $0.20$ & 10,11,12,13,14\\ 

FRB 20190417A    & $0.12817$  & $0.62$ & $5061$ & $190$ & $5$ & $0.62$ & 15,16\\ 

\hline  
   \end{tabular}
    {\flushleft
    \footnotesize{{\bf Notes:}\\  
$^{\mathrm{a}}$The measured redshifts of the host galaxies of FRBs. \\ 
$^{\mathrm{b}}$The luminosity distance calculated by redshift. The $\Lambda$CDM cosmological parameters are taken as $\Omega_{\rm m}=0.315$, $\Omega_bh^2=0.02237$, and $H_0=67.36\,{\rm km\,s^{-1}\,Mpc^{-1}}$ \citep{Planck20}. \\
$^{\mathrm{c}}$The observed RMs of FRBs. Here, we take the largest absolute value for the observed RMs.\\
$^{\mathrm{d}}$The observed PRS flux densities are primarily taken from measurements at 5–6 GHz, where data are available for all current five PRSs.\\
$^{\mathrm{e}}$To ensure consistency with the theoretical framework developed in this work, we adopt a uniform observed frequency of 5–6 GHz for all sources.\\
$^{\mathrm{f}}$The specific luminosities of the PRS inferred by the observed flux density and distance, i.e., $L_\nu=4\pi d_{\rm L}^2 F_\nu/(1+z)$.\\
$^{\mathrm{g}}$References: (1) \citet{Chatterjee17}, (2) \citet{Tendulkar17}, (3) \citet{Michilli18}, (4) \citet{Hilmarsson21}, (5) \citet{Niu22}, (6) \citet{Anna-Thomas23}, (7) \citet{Fong21}, (8) \citet{Xu22}, (9) \citet{Bruni24}, (10) \citet{Zhang25b}, (11) \citet{Bruni25}, (12) \citet{Zhang25}, (13) \citet{Bhusare25}, (14) \citet{Tian24}, (15) \citet{Moroianu26}, (16) \citet{Bruni26}.\\
}}
\end{table*}

Next, we propose that $\sigma_{\log R}$ can be measured by the scatter of the $L_\nu-{|\rm RM|}$ relation.
We consider the measured data as $(\log L_{\nu,i},\log {\rm |RM|}_i)$ with $i=1,2,3,...,N$, where $N$ is the number of the data. The data satisfy the potential physical relation of $L_{\nu,i}=K R_i^{\epsilon}|{\rm RM}|_i$, which can be equivalently expressed as
\begin{align}  
\log L_{\nu,i}=\log |{\rm RM}|_i+\epsilon\log R_i+\log K,\label{lumreal}
\end{align} 
where $R_i$ is the nebula radius as a random variable for each source.
According to Eqs.~(\ref{lum}), (\ref{zg2}), (\ref{lum2}) and the following sections, the particle momentum distribution $f(\hat p)$ is primarily determined by the nebula radius $R$. Therefore, the factor $\zeta_e\gamma_c^2$ mainly depends on $R$, which allows $K$ to be approximately regarded as a constant. Admittedly, the parameter $K$ implicitly absorbs other source-dependent physical quantities, such as magnetic field configuration, geometric factors of the nebulae, etc. Although these parameters vary from source to source and introduce an intrinsic scatter to $K$, this scatter is expected to be subdominant. Because the long-term dynamic evolution of FRB nebulae spans multiple orders of magnitude in radius $R$, the vast dynamic range of $R$ naturally dominates the observed global dispersion in the $L_\nu$--$|{\rm RM}|$ relation. Therefore, treating $K$ as an effective constant is physically well-justified within our current framework. Furthermore, as future surveys provide a larger sample size, the observed sources will populate an even broader range of evolutionary stages. This expanded dynamic range of $R$ will further marginalize the relative impact of the intrinsic scatter of $K$, reinforcing the robustness of treating it as a constant.
We first fit the observational data $(\log L_{\nu,i},\log {\rm |RM|}_i)$ using the following fitting function of 
\begin{align} 
\log L_{\nu,{\rm fit}}=\log {|\rm RM|}+C_0,\label{lumfit}
\end{align}  
where $C_0$ is the fitting intercept, ensuring the mean residual is zero, 
\begin{align}  
C_0=\left<\log L_{\nu,i}-\log|{\rm RM}|_i\right>
=\epsilon\left<\log R_i\right>+\log K.
\end{align}  
Notice that, regardless of whether RM is physically related to $R$ (for example, $|{\rm RM}|\propto R^{\delta}$), we deliberately enforce a positive correlation with a unit slope to fit the observational data, namely $L_\nu\propto |{\rm RM}|$. 
The detailed reasons for this choice will be explained in detail in the following Section \ref{validity}.
We then define the residual $\Delta$ for each data point as the difference between the observed value and the fitted value. Subtracting Eq.(\ref{lumfit}) from Eq.(\ref{lumreal}), $\Delta_i=\log L_{\nu,i}-\log L_{\nu,{\rm fit},i}$, the residuals simplify to
\begin{align} 
\Delta_i=\epsilon(\log R_i-\left<\log R_i\right>).
\end{align}  
Since the residuals must be centered at zero for any reasonable fitting algorithm, i.e., $\left<\Delta_i\right>=0$, the observed dispersion (standard deviation) of the residuals, $\sigma_{\Delta}$, is directly related to the standard deviation of $\log R$,
\begin{align} 
\sigma_{\Delta}=|\epsilon|\sigma_{\log R}.\label{sigmaDelta}
\end{align}  
Substituting Eq.(\ref{sigmalogR}) and Eq.(\ref{sigmaDelta}), we obtain the analytical expression connecting the observed scatter to $\alpha$ 
\begin{align} 
\alpha=(\ln 10)\fraction{\sigma_{\Delta}}{|\epsilon|}.\label{hatalpha}
\end{align}  
This result shows that the observed scatter directly measures the evolutionary index of the nebula.
Since $\log R$ has been considered to satisfy approximately an exponential distribution, The uncertainty of $\alpha$ is estimated based on the asymptotic variance of the maximum likelihood estimator for an exponential scale parameter, given by $\sigma_{\alpha}\simeq\alpha/\sqrt{N}$.
Based on the above result, in order to distinguish between various models with a difference of $\Delta\alpha$ larger than $3\sigma_{\alpha}$, the data number should satisfy 
\begin{align} 
N>9\fraction{\alpha}{\Delta\alpha}^2\sim\mathcal{O}(10).
\end{align} 
\emph{Therefore, we expect that a sample with $N\gg 10$ will be helpful to diagnose the physical origin of PRSs within $3\sigma$.}

\begin{figure}
    \centering
    \includegraphics[width = 1.0\linewidth, trim = 0 0 0 0, clip]{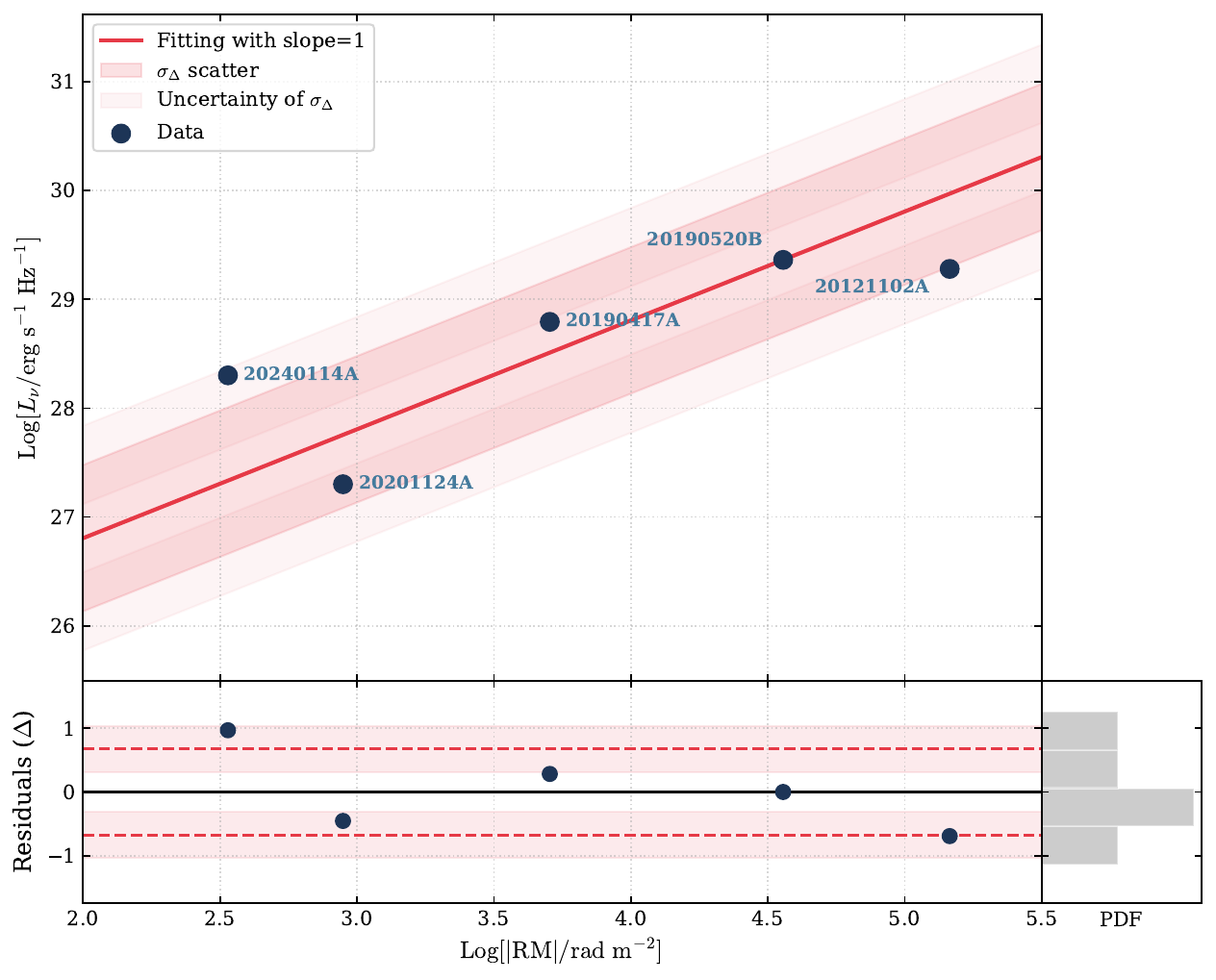} 
    \caption{The $L_\nu-|{\rm RM}|$ relation with the current five sources. Top panel: The observed data points are shown as black circles, labeled with their respective event IDs. The solid red line represents the best-fit linear model with a fixed slope of unity. The red shaded region indicates the $\sigma_\Delta$ scatter (standard deviation with its uncertainty) of the data around the best-fit line with slope equal to unity (See Section \ref{methodology} for a detailed justification of the fixed slope). 
    Bottom panels: The left panel displays the residuals ($\Delta$) of the fit in logarithmic scale. The horizontal red dashed lines mark the $\pm \sigma_\Delta$ boundaries (with the corresponding uncertainty labeled with red shaded region) estimated by Median Absolute Deviation (MAD) . The right panel shows the distribution of these residuals, derived from the sample statistics.
    }\label{relation}
\end{figure}

\subsection{Data Analysis of Observational Sample}

Despite the fact that we currently have only five confirmed PRSs, we still attempt to measure $\alpha|\epsilon|$ in this section, with detailed information shown in Table \ref{table1}.
Given that all current PRSs have observational data available in the 5–6 GHz band, we adopt the flux densities measured at this band. 
Notice that since the observational uncertainties in both PRS flux densities and FRB RMs are significantly smaller than the derived intrinsic scatter, the impact of measurement errors is negligible and thus omitted in our statistical framework.

Recognizing the limitations of the small sample size ($N=5$) of PRSs, we adopt a robust statistical framework to characterize the relationship between $L_\nu$ and $|\mathrm{RM}|$. 
For small-sample statistics where the Central Limit Theorem does not apply, we eschew traditional parametric estimators (based on the mean and standard deviation previously discussed in Section\ref{methodology}) in favor of robust alternatives based on the median and median absolute deviation (MAD). After a base-10 logarithmic transformation, we fit a linear model with a fixed unity slope, $\log L_{\nu,\mathrm{fit}} = \log |\mathrm{RM}| + C_0$, where the zero-point offset $C_0$ is determined by the median of the differences, $C_0 = \mathrm{Median}(\log L_\nu - \log |\mathrm{RM}|)$. The dispersion of the data is quantified using the MAD, a robust estimator of scale that is less sensitive to outliers and does not assume a specific distribution shape. To facilitate comparison with standard metrics, we compute the normalized MAD that corresponds to the standard deviation, defined as $\sigma_{\mathrm{MAD}} = 1.48\times\mathrm{Median}(|\Delta|)$, and take $\sigma_\Delta=\sigma_{\mathrm{MAD}}$ due to the limitations of the small sample size. Finally, we employ a non-parametric bootstrap resampling method ($N=10,000$) to estimate the 95\% confidence intervals for $\sigma_{\mathrm{MAD}}$, explicitly accounting for the large statistical uncertainties inherent in tiny datasets. 
Finally, we obtain
\begin{align}  
\sigma_\Delta=0.67\pm0.36.
\end{align}  
The fitting result and the standard deviation with its uncertainty are shown in Figure \ref{relation}.
According to Eq.(\ref{hatalpha}), this yields 
\begin{align} 
\alpha|\epsilon|=1.5\pm0.8.
\end{align} 
Due to the currently limited PRS sample size, our present constraints remain relatively weak. However, with larger statistical samples in the future, we will be able to provide much tighter constraints and unambiguously distinguish between various physical models of PRSs, as detailed in the Section \ref{sec4}.

\subsection{Mathematical Robustness}\label{validity}

A critical concern in our analysis is the potential unknown correlation between the independent variable RM and the stochastic variable $R$. Physically, these quantities might be coupled through an underlying function $|{\rm RM}| = g(R, X)$, where $X$ denotes other physical parameters like electron number density $n_e$ and magnetic field strength $B$ in ISM, which would imply that the intrinsic relationship between $\log L_\nu$ and $\log |{\rm RM}|$ may deviate from a linear function with a slope of unity. This raises the question of whether fitting a fixed-slope model ($\log L_\nu = \log |{\rm RM}| + C_0$) is mathematically justifiable and whether the derived $\alpha$ is biased by this assumption. 

Mathematically, the validity of our method does not rely on $|{\rm RM}|$ and $R$ being independent. The governing physical equation is defined as $L_\nu = K R^\epsilon |{\rm RM}|$. By fixing the slope to unity during the residual calculation, we perform a ``physical detrending'' rather than a statistical best-fit. The residual $\Delta$ is defined as $\Delta = \log L_\nu - \log |{\rm RM}| - C_0 = \epsilon \left<\log R_i\right> + \log k$.
This algebraic subtraction strictly isolates the contribution of $R$, regardless of whether $|{\rm RM}|$ itself is a function of $R$. Consequently, the observed residual $\Delta$ remains a direct proxy for $\log R$. If a correlation exists (e.g., $|{\rm RM}| \propto R^{\delta}$), it manifests as a systematic trend (tilt) in the residuals versus $\log |{\rm RM}|$. However, the global dispersion $\sigma_{\Delta}$ calculated over the entire sample correctly integrates both the intrinsic stochasticity of $R$ and the variation of $R$ induced by its coupling with RM. Thus, $\sigma_{\Delta}$ faithfully reflects the total dynamic range of $R$ within the observed sample.

\section{Constraints on Various Astrophysical Scenarios}\label{sec4}

Accumulating observational evidence suggests that the central engine of fast radio bursts (FRBs) is a neutron star, as exemplified by the Galactic event associated with SGR~J1935+2154 \citep{CHIME20,Bochenek20,Mereghetti20,Li21,Ridnaia21,Tavani21}. Motivated by this picture, several astrophysical scenarios can be considered for the origin of the magnetized environment responsible for both the observed RM and the associated PRS.

Broadly, these scenarios can be classified into the following categories:  
(1) \emph{Non-relativistic shock systems in nebulae}, including FS and RS in the SNR/ISM system, as well as FS formed in the PWN/SNR system, see the panel (a) of Figure \ref{shocks};  
(2) \emph{PWN bubble}, where relativistic fossil electron pairs are filling the region between the TS and the contact discontinuity in the PWN/SNR system, see the panel (a) of Figure \ref{shocks}; 
(3) \emph{Binary bow-shock systems}, in which a bow shock is formed through the interaction between the pulsar wind and the wind of a companion star, see the panel (b) of Figure \ref{shocks}.

\subsection{Non-relativistic shock in nebula}\label{shock} 

We first discuss the case of non-relativistic shocks in SNR and PWN, including FS/RS in the SNR/ISM system and FS in the PWN/SNR system, as shown in the panel (a) of Figure \ref{shocks}. The observed non-thermal radiation of PRSs implies that they are most likely to originate from relativistic electrons accelerated by shock. 
The thermal kinetic energy of electrons in the downstream of a non-relativistic shock wave is
\begin{align}  
E_{{\rm th}}\equiv k_BT=\frac{1}{2}\epsilon_em_pV_{\rm sh}^2,
\end{align} 
where $\epsilon_e$ is the proportion of thermal energy obtained by downstream electrons to the total available kinetic energy, $V_{\rm sh}$ is the shock velocity related to the upstream. 
We define the corresponding thermal momentum of electrons as $p_{\rm th}$.
In non-relativistic regime, $p_{\rm th}\simeq \sqrt{2m_eE_{\rm th}}$, leading to 
\begin{align}  
p_{\rm th}\sim(\epsilon_em_em_p)^{1/2}V_{\rm sh}.\label{pth}
\end{align}  

The next question is how many electrons are accelerated by the shock.
The electron injection problem in collisionless shocks arises from the scale disparity between the electron Larmor radius and the shock transition width, necessitating a pre-acceleration mechanism to elevate electrons from the thermal pool to a threshold momentum, $p_{\rm inj}$, required for diffusive shock acceleration (DSA). This threshold typically scales as a small multiple of the thermal momentum, 
\begin{align}  
p_{\rm inj}=\xi p_{\rm th},
\end{align} 
because it represents the optimal intersection between the super-thermal tail of the Maxwellian distribution and the kinematic requirement to outrun downstream advection. Below $p_{\rm inj}$, the electrons satisfy a thermalized Maxwellian distribution; and above $p_{\rm inj}$, the electrons satisfy a power-law distribution.
The independence of the ratio of $\xi\equiv p_{\rm inj}/p_{\rm th}$ from extrinsic parameters stems from the self-similar nature of shock microphysics: since both the downstream thermalization and the generation of mediating electromagnetic turbulence are self-consistently driven by the same reservoir of upstream ram pressure for strong shocks, the characteristic scales of the system scale proportionally, preserving the injection threshold as a universal constant of the normalized distribution. 

For non-relativistic shocks, most electrons in the downstream satisfy the Maxwell-Boltzmann distribution,
\begin{align}  
f(p)dp=\frac{4}{\sqrt{\pi}p_{\rm th}}\fraction{p}{p_{\rm th}}^2\exp\left[-\fraction{p}{p_{\rm th}}^2\right]dp,
\end{align}  
with $p_{\rm th}=\sqrt{2m_ek_BT}\propto V_{\rm sh}$ given by Eq.(\ref{pth}).
The peak momentum is $p_{\rm peak}\simeq p_{\rm th}$. 
We assume that a fraction of the electrons are accelerated above $\xi p_{\rm th}$ and the accelerated electrons satisfy a power-law distribution of
\begin{align} 
f(p)dp\propto \fraction{p}{p_{\rm inj}}^{-s}dp~~~\text{with}~p_{\rm inj}=\xi p_{\rm th}.
\end{align} 
Combining the above two distributions, one has
\begin{align} 
f(p)dp= f(z)dz\propto\left\{
\begin{aligned} 
&z^2e^{-z^2}dz, & z\ll \xi\\
&\xi^2e^{-\xi^2}\fraction{z}{\xi}^{-s}dz. & z\gg\xi
\end{aligned}\right.
\end{align} 
with $z\equiv p/p_{\rm th}$.
The CDF is
\begin{align} 
\zeta_e&\equiv F(>p_{\rm obs})=\fraction{z_{\rm obs}}{\xi}^{1-s}\nonumber\\
&\times\frac{4\xi^3}{\sqrt{\pi}(s-1)e^{\xi^2}{\rm erf}(\xi)+2\xi(2\xi^2-s+1)}\propto \fraction{p_{\rm obs}}{p_{\rm th}}^{1-s}\propto V_{\rm sh}^{s-1},\nonumber\\ 
\end{align}  
where $z_{\rm obs}\equiv p_{\rm obs}/p_{\rm th}$, and ${\rm erf}(...)$ is the Gauss error function. 
We assume that the relation between the shock radius and the shock velocity as
\begin{align} 
V_{\rm sh}\propto R^{\kappa_V}.
\end{align} 
According to the predicted $L_\nu-|{\rm RM}|$ relation and $\gamma_{c}\sim1$ in non-relativistic regime, one finally has
\begin{align} 
L\propto \zeta_e\gamma_c^2R^2|{\rm RM}|\propto R^{\epsilon} |{\rm RM}|~~~\text{with}~\epsilon=2+\kappa_V(s-1).\nonumber\\
\end{align}

\begin{table}
\centering
\caption{The theoretical values of the combined index $\alpha|\epsilon|$ for various shock scenarios compared with the measured value of $\alpha|\epsilon|=1.5\pm0.8$}\label{table_nebula}
\begin{tabular}{lccc}
\toprule
 & Phase & \multicolumn{2}{c}{$\alpha|\epsilon|$} \\
\cmidrule(lr){3-4}
 &  & $s=2.0$ & $s=3.0$ \\
\midrule
FS in SNR/ISM & Free-Expansion  & 2.0 & 2.0\\
FS in PWN/SNR & Free-Expansion  & 2.6 & 2.8\\
FS in SNR/ISM & Sedov-Taylor  & 0.2 & 0.4\\
RS in SNR/ISM & Free-Expansion  & 3.5 & 5.0\\
\bottomrule
\end{tabular} 
\end{table}

Since we have considered the shock radius evolve as $R\propto t^{\alpha}$, for the FSs in both the SNR/ISM system and the PWN/SNR system, the shock velocity evolves with $V_{\rm sh}\propto t^{\alpha-1}\propto R^{(\alpha-1)/\alpha}$, leading to $\kappa_V=(\alpha-1)/\alpha$. Therefore, 
1) in the free-expansion phase, the FS in the SNR/ISM system has $\alpha=1$ and $\kappa_V=0$, leading to $\epsilon=2$ and $\alpha|\epsilon|=2$; 
the FS in the PWN/SNR system has $\alpha=6/5$ and $\kappa_V=1/6$, leading to $\epsilon=2.2,2.3$ and $\alpha|\epsilon|=2.6,2.8$ for $s=2,3$, respectively.
2) in the Sedov-Taylor phase, the FS in the SNR/ISM system has $\alpha=2/5$ and $\kappa_V=-3/2$, leading to $\epsilon=0.5,-1$ and $\alpha|\epsilon|=0.2,0.4$ for $s=2,3$, respectively; the FS in the PWN/SNR system has been absent in this stage. 

For the RS in the SNR/ISM system, in the free-expansion phase, $R\propto t$ and $V_{\rm sh}\propto t^{3/2}$ (We should note that due to the contribution of an additional second-order small quantity \citep{McKee95}, a time-dependent shock velocity is obtained), one has $\alpha=1$ and $\kappa_V=3/2$, leading to $\epsilon=3.5,5.0$ and $\alpha|\epsilon|=3.5,5.0$ for $s=2,3$, respectively; in the Sedov-Taylor phase, the RS in the SNR/ISM system has been absent. All the results are summarized in Table \ref{table_nebula}. 
In conclusion, the measured value of $\alpha|\epsilon|=1.5\pm0.8$ disfavors SNRs in the Sedov–Taylor phase and reverse shocks during the free-expansion phase of SNR/ISM interactions, while being more consistent with forward shocks in the free-expansion phase of both SNR/ISM and PWN/SNR systems.

\subsection{PWN Bubble}\label{bubble}

In the above scenarios, the relativistic electrons with $\gamma\sim(10^2-10^3)$ that produce radio emission are accelerated by the shocks in SNR and ISM. However, in the TS of PWN, two different components of the relativistic electrons are to be distinguished \citep[e.g.,][]{Atoyan96,Gaensler06,Zhang08}, namely (a) the population of freshly accelerated ultrarelativistic electrons which are being presently injected downstream of the wind shock (so-called the wind electrons), and (b) the population of electrons of lower energies responsible for the radio emission, which are, presumably, `relic' and reflect the PWN history (so-called the fossil electrons). The wind electrons and the fossil electrons have a Lorentz factor of $\gamma\sim(10^4-10^6)$ and $\gamma\sim(10^2-10^3)$, respectively, which mainly produce the X-ray and radio emissions, respectively. The typical radio luminosity of PWNe ranges from $10^{28}~{\rm erg~s^{-1}}$ to $10^{36}~{\rm erg~s^{-1}}$ \citep{Frail97}.
In this work, since the PRSs are emitted at the radio band, we mainly focus on the fossil electrons. 
For the fossil electrons that emit the radio radiation, according to Eq.(\ref{gammaghz}), the synchrotron cooling timescale is estimated by
\begin{align} 
\tau_{\rm syn}=\frac{6\pi m_ec}{\sigma_{\rm T}B^2\gamma}\simeq4.1\times10^4~{\rm yr}\fraction{B}{1~{\rm mG}}^{-3/2}\fraction{\nu}{1~{\rm GHz}}^{-1/2},
\end{align}
which is most likely longer than the age of the nebula, meanwhile, the synchrotron cooling process can be ignored.
Due to the low energies and long radiative lifetime of fossil electrons, these particles are expected to be dominated by those previously injected into the PWN by the central neutron star, and therefore their spectrum reflects the evolution of the PWN during its history.
In this case, the fossil electrons are homogeneously distributed in the nebula with a radius at the contact discontinuity in the PWN/SNR system (shown in the panel (a) of Figure \ref{shocks}), and we term it the PWN bubble in this work, which also agrees with the diffuse radio maps of the Crab Nebula. Observationally, the fossil electrons are proposed to have a distribution of $n_e(\gamma)d\gamma\propto\gamma^{-s}$ with $s\sim(1-1.5)$ for $\gamma>\gamma_m$ \citep[e.g.,][]{Atoyan96,Gaensler06,Zhang08}.

The nebula bubble radius $R$ is determined by the evolutionary stage of the PWN/SNR system and the spindown of the neutron star \citep[e.g.,][also see Table \ref{table}]{Swaluw01}. In the free-expansion phase, the PWN bubble radius evolves as $R\propto t^{6/5}$. In the Sedov-Taylor phase, the bubble radius evolution determines whether the neutron star is in the spindown phase: 1) if the system age is smaller than the spindown timescale, the spindown luminosity is a constant, leading to a constant wind. In this case, the PWN bubble radius evolves as $R\propto t^{11/15}$; 2) if the system age is longer than the spindown timescale, the spindown luminosity decreases $t^{-2}$, leading to a decaying wind. In this case, the PWN bubble radius evolves as $R\propto t^{3/10}$. The derivation of the above scaling laws can be found in the Appendix \ref{appendix}.

In order to obtain how the factor $\zeta_e\gamma_c^2$ evolves with the bubble radius $R$, we first need to analyze the distribution evolution of the fossil electrons.
For the fossil electrons in a PWN bubble, the synchrotron cooling can be ignored and the dominant cooling process is adiabatic expansion.
Thus, the kinetic equation of electron distribution can be written as
\begin{align} 
\frac{\partial N_e(\gamma,t)}{\partial t}+\frac{\partial}{\partial\gamma}[\dot\gamma_{\rm ad}N_e(\gamma,t)]=Q(\gamma,t).
\end{align}
Assuming that the PWN expands adiabatically, one has the pressure $P\propto V^{-4/3}\propto R^{-4}$ for relativistic gas. The total internal energy is $U=N_e\gamma m_ec^2=3PV\propto R^{-1}$, finally leading to $\gamma\propto R^{-1}$ and $\dot\gamma_{\rm ad}=-(\dot R/R)\gamma$. Assuming $R\propto t^{\alpha}$ and defining $Q=(s-1)(\dot N_e/\gamma_m)(\gamma/\gamma_m)^{-s}$ for $\gamma>\gamma_m$ and $Q=0$ for $\gamma<\gamma_m$, the kinetic equation can be rewritten as
\begin{align} 
\frac{\partial N_e(\gamma,t)}{\partial t}-\frac{\alpha}{t}\frac{\partial}{\partial\gamma}[\gamma N_e(\gamma,t)]=(s-1)\frac{\dot N_{e}}{\gamma_m}\fraction{\gamma}{\gamma_m}^{-s}.
\end{align} 
It can be solved as  
\begin{align} 
N_e(\gamma,t)&=\frac{(s-1)}{1+\alpha(s-1)}\frac{\dot N_e t}{\gamma_m}\nonumber\\
&\times\left\{
\begin{aligned} 
&\fraction{\gamma}{\gamma_m}^{1/\alpha-1}, && \gamma_m(t/t_0)^{-\alpha}<\gamma<\gamma_m,\\
&\fraction{\gamma}{\gamma_m}^{-s}, && \gamma>\gamma_m.
\end{aligned}\right.
\end{align}
The electron number density is $n_e(\gamma,t)=N_e(\gamma,t)/V(t)\propto t^{1-3\alpha}$.  
The fossil electrons have an index of $s\sim1-1.5$ based on the PWN radio observation, thus, they should have a Lorentz factor $\gamma>\gamma_m$. Notice that the approximation of $t\gg t_0$ is used.
Therefore, the fraction of electron with $>\gamma$ is
\begin{align} 
\zeta_e\simeq\frac{(\gamma/\gamma_m)^{1-s}}{1+\alpha(s-1)}\sim\text{const. for }t\gg t_0.
\end{align} 
For the range of $\alpha$ considered here, and considering that the electron population is dominated by the low-energy band, Eq.~(\ref{gammac}) yields
\begin{align} 
\gamma_c\propto \left\{
\begin{aligned} 
&\gamma_m\fraction{t}{t_0}^{1/2-\alpha}\propto R^{1/2\alpha-1}, && \text{for }\alpha>1/2,\\
&\gamma_m\sim{\rm const}., && \text{for }\alpha<1/2.
\end{aligned}\right.
\end{align}
For the free-expansion phase and Sedov-Taylor phase with a constant wind, one always has $\alpha>1/2$, see Table \ref{table} in detail.
According to the predicted $L_\nu-|{\rm RM}|$ relation, one finally has
\begin{align} 
L\propto \zeta_e\gamma_c^2R^2|{\rm RM}|\propto R^{\epsilon} |{\rm RM}|~~~\text{with}~\epsilon= 1/\alpha.
\end{align}
This implies that $\alpha|\epsilon|= 1$, approximately consistent with the measured $\alpha|\epsilon|=1.5\pm0.8$.

On the other hand, the spindown scenario with a decaying wind, $\dot N_e=\dot N_{e,0}(t/\tau)^{-2}$, where $\tau$ is the spindown timescale, typically arises at the late stages of supernova evolution, e.g., in the Sedov-Taylor phase.
We assume $\alpha(s-1)<1$ for the range of parameters considered here, and $t\gg\tau$ and $t\gg t_0$, then one has
\begin{align} 
N_e(\gamma,t)&=\frac{(s-1)}{1-\alpha(s-1)}\frac{\dot N_{e,0} \tau}{\gamma_m}\fraction{\tau}{t_0}^{1-\alpha(s-1)}\fraction{t}{\tau}^{-\alpha(s-1)}\nonumber\\
&\times\fraction{\gamma}{\gamma_m}^{-s}~~~{\rm for}~\gamma>\gamma_m(t/t_0)^{-\alpha}.
\end{align}
Note that the rapid decay of the injection rate ensures that adiabatic cooling only shifts the injected power-law spectrum in energy without altering its slope, resulting in a single power-law fossil electron spectrum.
Besides, when taking into account an injection rate that decays with time as $t^{-2}$, the denominator $1-\alpha(s-1)$ in the above formula differs from that $1+\alpha(s-1)$ in the case of steady injection.
The fraction of electrons with Lorentz factor $>\gamma$ is
\begin{align} 
\zeta_e=\fraction{\gamma}{\gamma_m}^{1-s}\fraction{t}{t_0}^{-\alpha(s-1)}\propto R^{-(s-1)}.
\end{align}
Within the range of $s$ of interest, one has
\begin{align} 
\gamma_c\propto\gamma_m\fraction{t}{t_0}^{-\alpha}\propto R^{-1}.
\end{align}
For the Sedov-Taylor phase with a decaying wind, the temporal index of the nebula radius evolution is $\alpha=3/10$, see Table \ref{table} in detail.
According to the predicted $L_\nu-|{\rm RM}|$ relation, one finally has
\begin{align} 
L\propto \zeta_e\gamma_c^2R^2|{\rm RM}|\propto R^{\epsilon} |{\rm RM}|~~~\text{with}~\epsilon\sim-(s-1).
\end{align}
One finally has $\epsilon=0,-0.5$ and $\alpha|\epsilon|=0,0.15$ for the fossil electron with $s=1,1.5$, respectively. 
All the above results are summarized in Table \ref{table_bubble}.
In conclusion, the measured value of $\alpha|\epsilon|=1.5\pm0.8$ disfavors old PWNe powered by a decaying wind and is instead more consistent with young PWNe driven by a nearly constant wind.

\begin{table}
\centering
\caption{The theoretical values of the combined index $\alpha|\epsilon|$ for various PWN bubble scenarios compared with the measured value of $\alpha|\epsilon|=1.5\pm0.8$}\label{table_bubble}
\begin{tabular}{lccc}
\toprule
 & Phase  & \multicolumn{2}{c}{$\alpha|\epsilon|$} \\
\cmidrule(lr){3-4}
 & & $s=1.0$ & $s=1.5$ \\
\midrule
Constant Wind & Free-Expansion  & 1.0 & 1.0\\
Constant Wind & Sedov-Taylor  & 1.0 & 1.0\\
Decaying Wind & Sedov-Taylor  & 0.0 & 0.15\\
\bottomrule
\end{tabular} 
\end{table}

Although fossil electrons in a PWN bubble can naturally produce synchrotron radiation, a key issue concerns how the PWN contributes to the observable RM, since a relativistic pair plasma with a small amount of Goldreich–Julian net charge cannot generate a significant RM \citep{Yang23}. 
However, theoretical modeling of PWN dynamics and X-ray morphology suggests that baryons are expected to be present in PWNe, as their inclusion is physically required to address the so-called ``$\sigma$ problem'': a wind composed purely of electron–positron pairs cannot efficiently convert the pulsar's magnetic energy into the kinetic energy inferred at TS \citep{Kennel84}. Consequently, while electron–positron pairs dominate the particle number and the observable radiation, baryons are believed to carry a substantial fraction of the wind’s energy and to govern the large-scale dynamics of the nebula \citep{Hoshino92}, potentially augmented by external material mixed in from the SN ejecta \citep{Blondin01}. In this picture, the RM is expected to be contributed predominantly by the extra charge-neutralizing electrons associated with the baryonic component. 
Finally, because the PWN of a typical radio pulsar contributes a small value of RM, for a PRS to exhibit a significant RM in that region, the PWN would need to be atypical, for instance, a young PWN or a nebula dominated by magnetar winds \citep{Metzger19}.

\subsection{Bow shock in binary system}\label{binary}

In a binary system, the accelerated relativistic electrons generate non-thermal radiation at: 1) companion-wind shock; 2) pulsar-wind shock, as shown in the panel (b) of Figure \ref{shocks}.
First, for the pulsar-wind shock, unlike fossil electrons in a PWN/SNR system, which survive for millions of years and are confined by the surrounding ejecta, bow shocks in binary systems are, by contrast, open and compact: particles are rapidly advected out of the shock region on a much shorter timescales, preventing any long-term accumulation. In this case, only a small fraction of the pulsar wind power is intercepted by the shock, and only a minor portion of that energy is transferred to electrons radiating at radio frequencies.
On the other hand, for the companion-wind shock, since the shock velocity is only a few $\times100~{\rm km~s^{-1}}$, much less than that of the SNR shock, its synchrotron radiation would be much weaker. Due to the above reason, the radio luminosity of bow shocks is intrinsically low (e.g., $L\sim(10^{27}-10^{30})~{\rm erg~s^{-1}}$ with typical frequency of GHz, \citet{Dubus13,Deller15}), although the magnetic field near the bow shock can be strong, making individual electrons radiate efficiently.
As a result, the radio emission is governed by the instantaneous energy dissipation rather than by historical buildup, leading to radio luminosities several orders of magnitude lower than those of pulsar wind nebulae in supernova remnants.
If the PRSs are produced by the bow shock in a binary system, the binary should be much closer and the companion should have a stronger wind. In any case, we briefly discuss this situation below.

Unlike from the nebula scenario discussed in Section \ref{shock}, for the bow shock scenario in binary systems, the nebula radius mainly depends on the orbital separation and the pressure balance between the pulsar wind and the companion wind, rather than the nebula radius evolution. Thus, the scatter of the $L_\nu-|{\rm RM}|$ relation is determined by the binary parameter.
We assume that the orbital separations satisfy a power-law distribution
\begin{align} 
f(a)\propto a^{-\kappa_a},
\end{align} 
where $\kappa_a=1$ corresponds to the Öpik with $f(\log a)={\rm const}$. Based on pressure balance, the bow shock is located at a distance $R$ from the neutron star,
\begin{align} 
R=\frac{\sqrt{\eta_w}}{1+\sqrt{\eta_w}}a~~~\text{with}~\eta_w\equiv\frac{L_{\rm sd}}{\dot M_wv_{w}c},
\end{align}
where $L_{\rm sd}$ is the spindown luminosity of the neutron star, $\dot M_w$ is the mass loss rate of the companion, $v_{w}$ is the velocity of the companion wind. We assume that $\eta_w$ is independent of the orbital separation $a$, then the shock radius satisfies the distribution of 
\begin{align} 
f(R)\propto R^{-\kappa_a}.
\end{align}
On the other hand, the energy distribution of the accelerated electrons in both the shocked pair wind side and the shocked companion wind side mainly depends on the properties of the bow shock with parameter $L_{\rm sd},\dot M_w,v_w$. Both $\zeta_e$ and $\gamma_c$ are expected to be largely independent of $R$, as the bow-shock system is open, allowing cooling electrons to escape. Consider that the $L_\nu-|{\rm RM}|$ relation is more sensitive to $R$, the scatter of the relation might mainly determined by $f(R)\propto R^{-\kappa_a}$.
According to the predicted $L_\nu-|{\rm RM}|$ relation, one finally has 
\begin{align} 
L\propto \zeta_e\gamma_c^2R^2|{\rm RM}|\propto R^{\epsilon} |{\rm RM}|~~~\text{with}~\epsilon=2.
\end{align}
By combining with Eq.(\ref{CMFR}), we define $\tilde\alpha\equiv 1/(1-\kappa_a)$ and one finally has $\tilde\alpha|\epsilon|=2/(1-\kappa_a)$.
Notice that although the bow shock radius does not evolve with time, its statistical distribution among the population mathematically mimics the temporal evolution of an expanding nebula. Therefore, we can equate its spatial distribution index $\tilde\alpha$ to the effective temporal index $\alpha$ to utilize the same statistical framework.
By equating this spatial distribution index $\tilde\alpha$ to the effective variance parameter $\alpha$ derived from our scatter framework, one finally has $\kappa_a=-0.34\pm0.60$. 

The derived value has large uncertainties and is marginally consistent with a flat distribution ($\kappa_a\sim0$) or even Öpik's law ($\kappa_a\sim1$) within $2\sigma$. Therefore, the binary wind scenario cannot be ruled out solely based on this statistical argument. However, a more critical issue is the selection bias: radio luminosity is expected to drop significantly with separation $a$, meaning a flux-limited sample should be dominated by close binaries. 
Furthermore, we note that our derivation assumes a steady formation rate and an unbiased sampling of the PRS population. In reality, flux-limited surveys may suffer from Malmquist bias, preventing the detection of older, fainter, and larger nebulae.

\subsection{Co-existence of Expanding Nebula and Binary System}

While we have discussed the expanding nebula and the binary bow-shock scenarios independently, it is plausible that these two environments co-exist, particularly for actively repeating FRBs \citep{Wang25b}. In a realistic astrophysical picture, a young neutron star could reside in a binary system with a stellar companion while still being enveloped by a young, expanding nebula (e.g., a PWN inside an SNR). 

In such a hybrid scenario, the PRS luminosity and the RM may originate from distinct but physically coupled regions. The persistent radio luminosity $L_\nu$ would likely be dominated by the nebula (e.g., SNR or PWN). This is because the nebula traps particles over long timescales, leading to a much higher radiation efficiency compared to the compact and ``open'' binary bow-shock system, where particles are rapidly advected away. 

Conversely, the observed RM could receive complex contributions from both environments. The intense and highly variable RM components (including dramatic sign reversals, as seen in FRB 20190520B reported by \citet{Anna-Thomas23}) would naturally arise from the dynamic, turbulent stellar wind of the companion in the binary system. Meanwhile, the outer swept-up SNR ejecta could provide a substantial and relatively stable baseline RM contribution \citep{Yang23}. Consequently, the scatter in the $L_\nu - |{\rm RM}|$ relation for such hybrid systems would be a complex convolution of the nebula's temporal evolution and the binary orbital parameters. 

\section{Conclusion}\label{sec5}

Building upon the unified framework established by \citet{Yang20,Yang22}, we introduce in this work a novel statistical approach to diagnose the physical properties of these environments. 
Here, we demonstrate that the intrinsic scatter of this relation is not merely measurement noise but a valuable physical probe encoding the dynamical evolution of the nebula radius ($R \propto t^\alpha$). As future surveys increase the sample size, this method will enable direct constraints on the age distribution and evolutionary history of FRB environments. Ultimately, this approach provides a new pathway to probing the life cycle of magnetars and their surrounding nebulae.
By deriving a generic scaling of $L_\nu \propto R^\epsilon |\mathrm{RM}|$ and analyzing the residuals of the five currently confirmed FRB-PRS systems relative to the theoretical baseline, we successfully constrained the combined evolutionary index to be $\alpha|\epsilon| = 1.5 \pm 0.8$.

This new diagnostic method provides a robust framework to effectively discriminate between different astrophysical scenarios as larger population statistics become available. 
For example, SNRs in the Sedov–Taylor phase typically yield $\alpha|\epsilon| \sim 0.2-0.4$, and reverse shocks in SNR/ISM systems during the free-expansion stage are predicted to produce large values of $\alpha|\epsilon| \gtrsim 3.5$. Furthermore, PWNe powered by a decaying pulsar wind generally correspond to very small values of $\alpha|\epsilon| \sim 0-0.15$. The measured value of $\alpha|\epsilon| = 1.5 \pm 0.8$ therefore roughly falls outside the characteristic ranges associated with these scenarios. 
Instead, it lies broadly within or closer to the ranges expected for forward shocks propagating during the free-expansion phase, either in SNR/ISM systems or in PWN/SNR systems, where theoretical estimates give $\alpha|\epsilon| \sim 2.0-2.8$. It is also consistent with the expectations for young PWNe driven by an approximately constant pulsar wind, which typically predict $\alpha|\epsilon| \sim 1$.
These initial constraints tentatively favor the physical picture where active repeaters are powered by dynamically young, rapidly expanding nebulae (likely less than centuries old). 
While a binary bow-shock origin remains a mathematically plausible alternative depending on the orbital distribution, its inherently low radio luminosity implies that flux-limited samples would be heavily biased toward extremely close binaries, which contradicts the broad dispersion observed in the sample. Therefore, the current result more naturally supports the scenario of young, expanding SNRs or PWN bubbles, although it is still subject to considerable uncertainties.

We acknowledge that the current sample size ($N=5$) limits the statistical significance of these constraints, and future observations are needed to refine the evolutionary index. However, the scatter-analysis framework established here serves as a powerful, independent tool to map the evolutionary timeline of FRB sources. As next-generation radio surveys localize a larger population of repeating FRBs, this method will enable us to distinguish between various progenitor channels and trace the life cycle of the central engines by simply analyzing the dispersion in their $L_\nu - |\mathrm{RM}|$ phase space.

\section*{Acknowledgements} 
This independent work marks the 10th anniversary of the author's collaboration with Bing Zhang, originating from their early study on synchrotron-heated nebulae -- a theoretical exploration that predated the discovery of the first FRB-associated PRS.
The author thanks the anonymous referee for providing helpful comments and suggestions and also acknowledges the helpful discussions with Yue Wu, Chen-Hui Niu, Gabriele Bruni, Rui-Nan Li, and Xingyu Shao.
This work is supported by the National Natural Science Foundation of China (No. 12473047), the National Key Research and Development Program of China (No. 2024YFA1611603) and the Yunnan Key Laboratory of Survey Science (No. 202449CE340002). 

\section*{Data Availability}
This theoretical study did not generate any new data.

\bibliographystyle{mnras} 


\appendix

\section{Scaling Laws Presented in Table 1: Radius Evolution of Various Shocks and Bubbles in a PWN/SNR/ISM System}\label{appendix}

In this appendix, we briefly summarize the derivation of the scaling laws presented in Table \ref{table}. Detailed theoretical derivations can be found in \citet{McKee95} and \citet{Swaluw01}.

\subsection{SNR/ISM System}

\subsubsection*{1. Free-Expansion Phase}
At this early stage, the mass of the swept-up ISM is much less than the mass of the SN ejecta ($M_{\rm swept} \ll M_{\rm ej}$), meaning the ejecta experiences almost no deceleration and expands homologously.

(i) FS in the SNR/ISM interaction:
Because the ejecta is not significantly decelerated, the FS expansion velocity remains approximately constant, i.e., $V_{\rm sh} \sim {\rm const}$. Integrating this velocity over time yields a linear radius evolution, $R \propto t$.

(ii) RS in the SNR/ISM interaction:
The RS propagates inward in the ejecta's comoving frame, but it continues to expand outward in the observer's frame. 
For the unshocked medium in the ejecta interior, its absolute expansion velocity is proportional to its distance from the center, giving an upstream velocity ahead of the shock of $v_{\rm up} = R/t$. While the mass of the inner ejecta core is constant, its volume expands, causing the density to drop sharply with time as $\rho_{\rm ej} \propto t^{-3}$.

The RS is driven by the pressure exerted by the external ISM. During this phase, the outermost FS sweeps through a uniform ISM at a constant velocity, resulting in a constant post-shock pressure, $P_{\rm FS} \sim {\rm const}$. This inward thermal pressure $P_{\rm FS}$ balances the ram pressure $P_{\rm RS}$ generated by the SN ejecta crashing into the RS. According to shock jump conditions, the ram pressure at the RS is $P_{\rm RS} \propto \rho_{\rm ej} V_{\rm sh}^2$, where $V_{\rm sh}$ is the velocity of the RS relative to the unshocked ejecta. Substituting $\rho_{\rm ej} \propto t^{-3}$ into the balance equation ($P_{\rm FS} \sim P_{\rm RS}$), one obtains:
\begin{align}
    V_{\rm sh} \propto t^{3/2}.
\end{align}
Kinematically, $V_{\rm sh}$ is the difference between the outward velocity of the upstream medium ($v_{\rm up} = R/t$) and the outward propagation velocity of the RS in the observer's frame ($v_{\rm shock} = dR/dt$):
\begin{align}
    V_{\rm sh} = \left| v_{\rm up} - v_{\rm shock} \right| = \frac{R}{t} - \frac{dR}{dt}.
\end{align}
Solving this differential equation for the RS radius yields:
\begin{align}
    R(t) = At - Bt^{5/2},
\end{align}
where $A$ and $B$ are constants. In the free-expansion phase (where the timescale is much less than the Sedov-Taylor timescale), the first-order linear term is dominant, leading to the macroscopic observation that $R \propto t$.

\subsubsection*{2. Sedov-Taylor Phase}
The swept-up ISM mass now greatly exceeds the ejecta mass ($M_{\rm swept} \gg M_{\rm ej}$), and the system enters an energy-conserving adiabatic expansion phase.

(i) FS in the SNR/ISM interaction: 
Since the physical processes during this stage are determined solely by the total SN explosion energy, $E_{\rm SNR}$, and the uniform mass density of the unshocked ISM, $\rho_{\rm ISM}$, the time-evolution of the FS radius can be derived strictly through dimensional analysis:
\begin{align}
    R \propto E_{\rm SNR}^{1/5}\rho_{\rm ISM}^{-1/5}t^{2/5} \propto t^{2/5}.
\end{align}
Taking the derivative with respect to time gives the FS velocity, $V_{\rm sh} = dR/dt \propto t^{-3/5}$. 

(ii) RS in the SNR/ISM interaction:
By this stage, the reverse shock has already propagated to the center of the ejecta and vanished.

\subsection{PWN Bubble}

The expansion of the PWN bubble is governed by the energy injected by the central neutron star and the confining pressure of the surrounding SNR. Note that the outer boundary of the PWN bubble corresponds to a contact discontinuity rather than a shock front. Furthermore, in this scenario, the radio emission is primarily produced by fossil electrons trapped within the bubble.

\subsubsection*{1. Free-Expansion Phase}
In this stage, the neutron star is dynamically young, spin-down is negligible, and the wind luminosity remains constant ($L \sim {\rm const}$). The PWN expands supersonically into the cold, freely expanding SN ejecta ($\rho_{\rm ej} \propto t^{-3}$). The total energy injected into the PWN is $E_{\rm PWN} \sim Lt$, producing an internal bubble pressure of $P_{\rm PWN} \sim E_{\rm PWN}/R^3 \sim Lt/R^3$. This internal pressure is balanced by the ram pressure of the ejecta, $P_{\rm SNR} \sim \rho_{\rm ej}(dR/dt)^2 \propto R^2/t^5$. Equating the two pressures yields $R \propto t^{6/5}$.

\subsubsection*{2. Sedov-Taylor Phase (Constant Wind)}
Once the SNR enters the Sedov-Taylor phase, its radius evolves as $R_{\rm SNR} \propto t^{2/5}$ as pointed out above. The RS has already crushed the ejecta, heating it significantly. The confining pressure is now the thermal pressure inside the SNR, governed by $P_{\rm SNR} \sim E_{\rm SNR}/R_{\rm SNR}^3 \propto t^{-6/5}$ assuming $E_{\rm SNR} \sim {\rm const}$. Assuming the pulsar wind is still constant ($L \sim {\rm const}$), the PWN pressure remains $P_{\rm PWN} \sim Lt/R^3$. The PWN now expands subsonically, driven by the thermal pressure balance $P_{\rm PWN} \sim P_{\rm SNR}$, which leads to $R \propto t^{11/15}$.

\subsubsection*{3. Sedov-Taylor Phase (Decaying Wind)}
As the system's age exceeds the pulsar's spin-down timescale, the wind luminosity drops sharply as $L \propto t^{-2}$. Consequently, the energy injected during the early stages dominates, and the total energy no longer increases linearly. The bubble's energy is now primarily depleted via adiabatic expansion. For a PWN bubble filled with relativistic gas, the pressure and volume obey the adiabatic relation $P_{\rm PWN} \propto V^{-4/3}$. 
The internal energy during this adiabatic expansion scales as $E_{\rm PWN} \propto P_{\rm PWN}V \propto V^{-1/3} \propto R^{-1}$. Therefore, the internal pressure of the PWN scales as $P_{\rm PWN} \sim E_{\rm PWN}/R^3 \propto R^{-4}$. Equating this to the confining Sedov pressure of the SNR ($P_{\rm SNR} \propto t^{-6/5}$) yields $R \propto t^{3/10}$.

\subsection{Forward Shock in the PWN/SNR System}

This FS is generated ahead of the PWN bubble as long as it pushes supersonically into the unshocked SNR ejecta.

\subsubsection*{1. Free-Expansion Phase}
During this phase, the FS closely trails the contact discontinuity (i.e., the PWN bubble). Therefore, its radius evolves identically to the bubble, $R \propto t^{6/5}$. The relative shock velocity $V_{\rm sh}$ depends on the difference between the absolute velocity of the shock ($v_{\rm shock} = dR/dt$) and the outward velocity of the upstream medium ($v_{\rm up} = R/t$) in the observer's frame:
\begin{align}
    V_{\rm sh} = \left| v_{\rm shock} - v_{\rm up} \right| = \frac{dR}{dt} - \frac{R}{t} \propto t^{1/5}.
\end{align}

\subsubsection*{2. Sedov-Taylor Phase}
When the outer SNR enters the Sedov-Taylor phase, the SNR's RS has already swept through and drastically reheated the entire ejecta. This extreme heating causes the sound speed of the ambient gas to rise significantly. As a result, the expansion velocity of the PWN bubble becomes subsonic relative to the local sound speed. Because the fluid motion is no longer supersonic, this inner FS simply cannot form.

\bsp    
\label{lastpage} 

\end{document}